\documentstyle[aps,preprint]{revtex}
\begin{document}
\draft
\title{ The Speed of Fronts of the Reaction-Difusion Equation}
\author{R.\ D.\ Benguria and M.\ C.\ Depassier}
\address{        Facultad de F\'\i sica\\
	P. Universidad Cat\'olica de Chile\\ Casilla 306,
Santiago 22, Chile}

\maketitle
\begin{abstract} We study the speed of propagation of
fronts for the scalar reaction-diffusion equation $u_t =
u_{xx} + f(u)$\, with $f(0) = f(1) = 0$.  We give a new
integral variational principle for the speed of the fronts
joining the state $u=1$ to $u=0$. No assumptions are made
on the reaction term $f(u)$ other than those needed to
guarantee the existence of the front.  Therefore our
results apply to the classical case $f > 0$ in $(0,1)$, to
the bistable case and to cases in which $f$ has more than
one internal zero in $(0,1)$.
\end{abstract}


The one dimensional reaction diffusion equation
\begin{equation}
u_t = u_{xx} + f(u) \qquad {\rm with}\qquad  f(0) = f(1) = 0
\label{eq:u}
\end{equation}
with $f(u) \in C^1[0,1]$
has been the subject of much study as it models diverse
phenomena in biology, population dynamics, chemical
physics, combustion and others
\cite{Murray,Britton,Showalter,Volpert,Pikelner}. Not only
is it itself of interest, but based on the rigourous
results available for this equation, diverse methods
applicable to pattern forming systems have been developed
\cite{Langer,Benjacob,VanSaarloos,Goldenfeld}.  In
applications, the reaction term $f(u)$ obeys additional
requirements  depending on the phenomenon being modelled.
Three types of nonlinearities appear to be generic, and are
shown in Fig. 1. Type A, for which $f > 0$ in
$(0,1)$ is the class to which the classical case of
Fisher\cite{Fisher} and Kolmogorov, Petrovskii and
Piskounov (KPP) \cite{KPP37} belongs. Type B, usually
referred to as the combustion case satisfies $f=0$ on
$(0,a)$ and $f >0$ on $(a,1)$, while finally, type C,
called the bistable case, satisfies $f(u) < 0$ for $u$ in
$(0,a)$, $f >0$ on $(a,1)$ with $\int_0^1 f(u)\, du >0$. More
general cases for $f$, namely cases in which $f$ has more
than one internal zero have also been studied
\cite{Fife1,Fife2}.

The time evolution of an initial condition $u(x,0)$ has
been studied for all the cases mentioned. It was proved
\cite{AW78} that for suitable initial conditions the
disturbance evolves into a monotonic travelling front $u =
q(x-c\,t)$ joining the stable state $u=1$ to $u=0$. In case
A there is continuum of values of $c$ for which a monotonic
front exists, the system evolves into the front of minimal
speed.  In cases B and C there is a single isolated value
of the speed for which the front exists. In these two last
cases there are threshold effects, necessary conditions for
the evolution of the system into the front have been
established as well \cite{Fife1,AW78}.  The same is true
for reaction terms with more than one internal
zero\cite{Fife1}. The problem which interests us here is
the determination of the asymptotic speed of the front.
There have been numerous studies of this problem,  a very
complete review is given in \cite{Volpert}. For reaction
terms of type A which in addition satisfy $f'(0) > f(u)/u$
the speed is given by \cite{KPP37}\,$c = c_{KPP} = 2
\sqrt{f'(0)}$.  Another reaction term of type A, that of a
function $f$ approaching a Dirac function at $u=1$ was
studied related to combustion by Zeldovich and
Frank--Kamenetskii.  They showed \cite{ZFK38} that the
speed tends to $c_{ZFK} = \sqrt{2 \int_0^1 f(u)\, du}$. For an
arbitrary reaction function $f(u)$ a local variational principle
of the
minimax type exists \cite{Volpert,HR75}. For reaction terms
of type A and B we have shown that  an integral variational
principle of the Rayleigh Ritz type exists \cite{BD1,BD2}.
Recently an interesting conjecture\cite{Gor95} has been put
forward for a restricted class of reaction functions.

The purpose of this article to show that the speed of the
front joining the state $u=1$ to $u=0$ derives from an
integral variational principle without any restriction on
$f$ other than those needed to guarantee the existence of
the front. The derivation follows an  approach similar to
the one used to obtain  the principle  valid only for for
positive reaction terms. However, this new principle which
is  valid for all cases, is not related, nor equivalent,
to the previous
one.

It is known\cite{AW78} that for a function of type A, B, or
C there exists a strictly decreasing front $u = q(x- c t)$
joining $u=1$ to $u=0$ for some $c >0$.  The front satisfies
$q_{zz} + c q_z + f(q) = 0$, $\lim u_{z \rightarrow
-\infty} = 1$, $\lim u_{z \rightarrow \infty} = 0$, where
$z = x - c t$.  Following the usual procedure, since the
front is monotonic, we define $p(q) = - dq/dz$, where the
minus sign is included so that $p$ is positive. One finds
that the monotonic fronts are  solutions of
\begin{mathletters}
\begin{equation}
p(q)\, {dp\over dq} - c \, p(q) + f(q) = 0,
\label{eq:p}
\end{equation}
with
\begin{equation} p(0) = 0, \qquad p(1) = 0, \qquad p > 0
\quad  {\rm in}\quad (0,1).
\end{equation}
\end{mathletters}
The derivation follows in a simple way from equation
(\ref{eq:p}).  Let $g$ be any positive function in (0,1)
such that $h = - dg/dq > 0$.  Multiplying equation
(\ref{eq:p}) by $g(q)$ and integrating between $q = 0$ and
$q =1$ we obtain, after integration by parts,
\begin{equation}
 \int_0^1 f\, g\, dq =   c \int_0^1 p\, g\, dq - \int_0^1
{1\over 2} h\, p^2\, dq.
\end{equation}
However, since $p$, $g$ and $h$ are positive, for fixed
$q$, the function $$
\phi(p) = c\, p\, g - {1\over 2} h\, p^2
$$
has a maximum at
\begin{equation}
p_{max} = c {g\over h}
\label{eq:pmax}
\end{equation}
so
$$
\phi(p) \le c^2 {g^2\over 2 h}
$$
at each value of $q$. It follows then that
\begin{equation}
c^2\, \ge 2\, {{\int_0^1  f\, g\, dq}\over{\int_0^1
 (g^2/h) dq}}.
\label{eq:bound}
\end{equation}
This bound on the speed is valid for any $f$ for which a
monotonic front exists. To show that this is a variational
principle we must show that there exists a function ${\hat
g}$ at which the equality holds. From Eq. (\ref{eq:pmax})
we see that $$ c {{\hat g}\over{\hat h}} = p $$ which can
be integrated. The maximizing $g$ is given by
\begin{equation}
{\hat g} = \exp \left( - \int_{q_0}^q {c\over p}\, dq \right)
\label{eq:gsol}
\end{equation}
with $0 < q_0 < 1$.
Evidently ${\hat g}$ is positive, monotonic decreasing and
moreover ${\hat g} (1) = 0$. Near $q = 0$ ${\hat g}$
diverges. We must ensure that the integrals in Eq.
(\ref{eq:bound}) exist. To verify this we recall
\cite{AW78} that in the three cases, A, B and C, the front
approaches $q=0$ exponentially, therefore, near zero, $$ p
\sim  {1\over 2} \left( c + \sqrt{c^2 - 4 f'(0)}\right)\,
q\, \equiv m\, q.  $$ Then we obtain $$ {\hat g}(q) \sim
{1\over q^{c/m}}$$ near zero and $f{\hat g}$ and ${\hat
g}^2/{\hat h}$ diverge at most as $ q^{1- (c/m)}$. The
integrals in Eq.(\ref{eq:bound}) exist if $m/c > 1/2$. This
condition is always satisfied when $f'(0) < 0$, that is in
cases B and C. In case A this condition is satisfied
provided that $c > 2\sqrt{f'(0)}$, which is the KPP case.
However, choosing as a trial function $g(q) = \alpha (2 -
\alpha ) u^{\alpha -2}$ with $0 < \alpha < 1$ one can check
that in the limit $\alpha \rightarrow 0$, $c^2 \rightarrow
4 f'(0)$.  Therefore, including all cases our main result
is
\begin{equation}
c^2\,=  \sup \left( 2\, {{\int_0^1  f\, g\, dq}\over{\int_0^1
 (-g^2/g') dq}} \right),
\label{eq:var}
\end{equation}
where the supremum is taken over all positive decreasing
functions $g$
in $(0,1)$ for which the integrals exist.
Moreover there is always a maximizing $g$ in cases B and C,
 while for
case A
there is a maximizing $g$ whenever $ c > c_{KPP}$.

While we considered the case of decreasing fronts, similar
results can be derived for increasing fronts and for the
case of density dependent diffusion following the same
approach.

As an example we may apply the above result to the Nagumo
equation which corresponds to a reaction term of the form
$$ f(u) = u (1-u) (u - a) \qquad {\rm with} \qquad 0 < a
<1/2 $$ This reaction term is of the bistable type. For
this equation the solution to Eq.(\ref{eq:p}) is known, it
is given by $p(q) = {1\over \sqrt{2}} q (1 - q)$ and the
speed is given by $$ c= {1\over \sqrt{2}} - a \sqrt{2}.  $$
To exhibit in this solvable case that the exact speed can
be obtained from the variational principle choose as a
trial function $$ g(q) = \left( {{1-q}\over q} \right)^{1 -
2 a} $$ The integrals can be performed easily. We obtain
$$
\int_0^1 (-g^2/g') dq = {{\Gamma (1+2 a)
\Gamma (3 - 2 a)}\over {(1 - 2 a)\Gamma (4)}}
$$
and
$$
 \int_0^1 f\, g\, dq = {{(1 - 2 a)\Gamma (1+2 a)
 \Gamma (3 - 2 a)}\over {4 \Gamma (4)}}
$$
so that $c^2 = (1 - 2 a)^2/2$ which is the exact value.
For other non solvable cases, it is a simple matter to
obtain accurate values for the speed using standard
variational techniques.

\section{Acknowledgments}

This work has been partially supported by Fondecyt project 1930559.

\end{document}